\documentclass[12pt]{article}
\usepackage{latexsym}
\usepackage{amssymb}
\usepackage{graphicx}
\usepackage{cite}

\newtheorem{Theorem}{Theorem}[part]
\newtheorem{Definition}{Definition}[part]
\newtheorem{Proposition}{Proposition}[part]

\newtheorem{Lemma}{Lemma}[part]

\newtheorem{Example}{Example}[part]

\def \ep{\hbox{ }\hfill$\Box$}

\def\reff#1{{\rm(\ref{#1})}}

\begin{document}
\title{Geometric measure of entanglement of multipartite mixed states}

\author{
Shenglong Hu \thanks{Email: Tim.Hu@connect.polyu.hk. Department of
Applied Mathematics, The Hong Kong Polytechnic University, Hung Hom,
Kowloon, Hong Kong. This author's work was partially supported by the National Natural Science Foundation of China (Grant
No. 11101303).},\hspace{6mm}Liqun Qi \thanks{Email:
maqilq@polyu.edu.hk. Department of Applied Mathematics, The Hong
Kong Polytechnic University, Hung Hom, Kowloon, Hong Kong. This
author's work was supported by the Hong Kong Research Grant
Council.},\hspace{6mm}Yisheng Song \thanks{Email: songyisheng123@yahoo.com.cn. Department of
Applied Mathematics, The Hong Kong Polytechnic University, Hung Hom,
Kowloon, Hong Kong.},\hspace{6mm} Guofeng Zhang \thanks{Email:
magzhang@inet.polyu.edu.hk. Department of Applied Mathematics, The
Hong Kong Polytechnic University, Hung Hom, Kowloon, Hong Kong. This
author's work was supported by the Hong Kong Research Grant
Council.}}

\date{\today} \maketitle

\begin{abstract}
\noindent The geometric measure of entanglement of a pure state,
defined by its distance to the set of pure separable states, is
extended to multipartite mixed states. We characterize the nearest
disentangled mixed state to a given mixed state with respect to this
measure by a system of equations. The entanglement eigenvalue for a
mixed state is introduced.  For a given mixed state, we show that
its nearest disentangled mixed state is associated with its
entanglement eigenvalue.

\vspace{3mm}

\noindent {\bf Key words:}\hspace{2mm} quantum entanglement, geometric measure
\vspace{3mm}
\end{abstract}


\section{Introduction}

\setcounter{Assumption}{0} \setcounter{Theorem}{0}
\setcounter{Proposition}{0} \setcounter{Corollary}{0}
\setcounter{Lemma}{0} \setcounter{Definition}{0}
\setcounter{Remark}{0} \setcounter{Algorithm}{0}
\setcounter{Example}{0} \hspace{4mm} The quantum entanglement
problem is regarded as a central problem in quantum information
\cite{nc00,v00,pv07}, and the geometric measure is one of the most
important measures of quantum entanglement
\cite{bh01,s95,wg03,pv07}. It was first proposed by Shimony
\cite{s95} and generalized to multipartite systems by Wei and
Goldbart \cite{wg03}, and has become one of the widely used
entanglement measures for multiparticle cases
\cite{hkwg09,hmmov09,cxz10,hs10, q122,gbb08,hqz12}.

The geometric measure is based on the geometric distance between a given pure state and the set of separable pure states. From the definition, the quantum eigenvalue problem is derived to characterize the nearest separable pure state with respect to this measure \cite{wg03,hs10,q122}. This characterization is significant: {\bf the eigenvalues are always real numbers and the largest one corresponds to the maximal overlap of the given pure state and the separable pure states}.

Based on the convex roof construction, this geometric measure is
extended to multipartite mixed states \cite{wg03}. Although the
extension is standard, analogue characterizations for disentangled
mixed states are not clear \cite{wg03,hkwg09}. Instead of the convex
roof extension, we propose in this paper a natural extension of the
geometric measure from pure states to mixed states. Most
interestingly, a characterization for the nearest disentangled mixed
state still holds. We show that there is a system of equations
associated to the proposed geometric measure for mixed states. The
entanglement eigenvalue for a mixed state is introduced and it is
proven to be an indicator of the proposed geometric measure.
Moreover, the disentangled mixed state corresponding to the
entanglement eigenvalue is shown to be the nearest disentangled
mixed state to the given mixed state with respect to this measure.

The rest of this paper is organized as follows. Some preliminaries are presented in Section 2 to include some basic definitions. The geometric measure of mixed states is proposed in Section 3. In Section 4, the characterization for the nearest disentangled mixed state is investigated. Section 5 concludes this paper with some remarks.


\section{Preliminaries}

\setcounter{Assumption}{0}
\setcounter{Theorem}{0} \setcounter{Proposition}{0}
\setcounter{Corollary}{0} \setcounter{Lemma}{0}
\setcounter{Definition}{0} \setcounter{Remark}{0}
\setcounter{Algorithm}{0}  \setcounter{Example}{0}

\hspace{4mm} An $m$-partite pure state $|\Psi\rangle$ of a composite
quantum system can be regarded as a normalized element in a Hilbert
tensor product space
${\mathcal{H}}=\bigotimes_{k=1}^m{\mathcal{H}}_k$, where the
dimension of ${\mathcal{H}}_k$ is $d_k$ for $k=1,\ldots,m$. A
separable $m$-partite pure state $|\Phi\rangle\in{\mathcal{H}}$ can
be described by $|\Phi\rangle=\bigotimes_{k=1}^m|\phi^{(k)}\rangle$
with $|\phi^{(k)}\rangle\in{\mathcal{H}}_k$ and
$\||\phi^{(k)}\rangle\|=1$ for $k=1,\ldots,m$. Denote by
$Separ(\mathcal{H})$ the set of all separable pure states in
$\mathcal H$.  A state is called entangled if it is not separable.

For a given $m$-partite pure state $|\Psi\rangle\in {\mathcal{H}}$,
a geometric measure is then defined as \cite{wg03}
\begin{eqnarray} \label{gm1}
\min_{|\Phi\rangle \in Separ(\mathcal{H})}\| | \Psi \rangle - | \Phi
\rangle \|,
\end{eqnarray}
or one may consider
\begin{eqnarray}\label{gm2}
\min_{|\Phi\rangle \in Separ(\mathcal{H})}\frac{1}{2}\| | \Psi \rangle - | \Phi
\rangle \|^2 = 1-G(\Psi),
\end{eqnarray}
where $G(\Psi)$ is the maximal overlap:
\begin{eqnarray}\label{me}
G(\Psi)=\max_{|\Phi\rangle \in
Separ(\mathcal{H})}|\langle\Psi|\Phi\rangle|.
\end{eqnarray}

Based on \reff{gm2}, the quantum eigenvalue problem is proposed and analyzed in
\cite{wg03,q122}:
\begin{eqnarray}\label{qe}
\left\{\begin{array}{rcl}\langle \Psi|\left(\bigotimes_{j\neq k}|\phi^{(j)}\rangle\right)&=&\lambda\langle\phi^{(k)}|,\\\left(\bigotimes_{j\neq k}\langle\phi^{(j)}|\right)\Psi\rangle&=&\lambda|\phi^{(k)}\rangle,\\\||\phi^{(k)}\rangle\|&=&1,k=1,\ldots,m.\end{array}\right.
\end{eqnarray}
\begin{Proposition}\label{prop-2}
Let $|\Psi\rangle\in{\cal H}$ be a pure state and the corresponding quantum eigenvalue problem be \reff{qe}.
Then, $\lambda$ is a real number and the maximal overlap in
\reff{me} is equal to the largest such $\lambda$.
\end{Proposition}

\noindent {\bf Proof.} See \cite[Section II]{wg03} or \cite[Section 2]{q122} for the detailed proof. \ep

The largest $\lambda$ in \reff{qe}, denoted by $\Lambda_{\max}$, is called the entanglement eigenvalue \cite{wg03,q122}. Consequently, the geometric measure in \reff{gm2} equals $1-\Lambda_{\max}$.

The entanglement problem for mixed states in ${\cal H}$ has
attracted much attention as well
\cite{wg03,hkwg09,pv07,gbb08,w98,stv98}. Usually, a mixed state in
${\cal H}$ is represented by a density matrix $\varrho$ of size
$\prod_{k=1}^md_k\times \prod_{k=1}^md_k$ \cite{hkwg09,wg03,x96}. It
is Hermitian, positive semidefinite and trace one. There are several
concepts on disentangled multipartite mixed states. We adopt the
following one \cite{pv07}.

\begin{Definition}\label{def-1}
For a mixed state in ${\cal H}$ with density matrix $\varrho$, it is disentangled if
\begin{eqnarray*}
\varrho=\sum_{k}p_k|\Psi^{(k)}\rangle\langle\Psi^{(k)}|
\end{eqnarray*}
for some pure separable states $|\Psi^{(k)}\rangle\in Separ({\cal H})$,
$p_k\geq 0$ and $\sum_{k}p_k=1$.
\end{Definition}

Denote by
$Disen({\cal H})$ the set of all disentangled mixed states in ${\cal H}$.

The geometric measure for pure states can be extended to mixed states through the convex roof construction \cite{wg03}:
\begin{eqnarray}\label{gm5}
E_C(\varrho):=\min_{\{p_i,\Psi^{(i)}\}}\sum_{i}p_iM(|\Psi^{(i)}\rangle)
\end{eqnarray}
where the minimum is taken over all decompositions $\varrho=\sum_{i}p_i|\Psi^{(i)}\rangle\langle\Psi^{(i)}|$ into pure states with the $p_i$ forming a probability distribution, and the measure $M$ for pure states can be chosen to be both the measure \reff{gm2} and any other measures.


\section{Geometric measure for mixed states}

\setcounter{Assumption}{0} \setcounter{Theorem}{0}
\setcounter{Proposition}{0} \setcounter{Corollary}{0}
\setcounter{Lemma}{0} \setcounter{Definition}{0}
\setcounter{Remark}{0} \setcounter{Algorithm}{0}
\setcounter{Example}{0} \hspace{4mm} Although the geometric measure
defined in \reff{gm5} satisfies the criteria for entanglement
monotone \cite{v00,wg03}, the extension of Proposition \ref{prop-2}
to mixed states is not clear and there lack characterizations of the
nearest disentanglement mixed state to an arbitrary mixed state. In
this section, instead of \reff{gm5}, we propose a geometric measure
for mixed states which is a natural extension of \reff{gm2}.

\begin{Definition}\label{def-3}
For a mixed state in ${\cal H}$ with density matrix $\varrho$, its geometric measure is defined as:
\begin{eqnarray}\label{gm-m}
E(\varrho):=\min_{\rho\in Disen({\cal
H}),\;\|\rho\|=\|\varrho\|}\|\varrho-\rho\|,
\end{eqnarray}
where the norm is the Frobenius norm of matrices.
\end{Definition}

To see that Definition \ref{def-3} is well-defined, the following lemma is essential.

\begin{Lemma}\label{lem-1}
Let $\varrho$ be the density matrix of a mixed state in ${\cal H}$. Then the set $S(\varrho):=\{\rho\in Disen({\cal H})\;|\;\|\rho\|=\|\varrho\|\}$ is a nonempty compact set.
\end{Lemma}

\noindent {\bf Proof.} Let $n:=\Pi_{k=1}^md_k$ and $N:=n^2+1$. The
density matrix is an $n\times n$ Hermitian matrix. For any
density matrix $\varrho$, its real part is a symmetric $n\times n$
matrix and its imaginary part is an skew-symmetric $n\times n$
matrix. Consequently, the real dimension of $Disen({\cal H})$ is $n^2$. By
the definition of $Disen({\cal H})$, every $\rho\in Disen({\cal H})$
can be represented as a convex combination of density matrices of
pure separable states. By Caratheodory's theorem\cite{r70}, the
number of density matrices of pure separable states in such a
combination can be chosen to be at most $N$. Consequently, we have
\begin{eqnarray*}
S(\varrho)=\left\{\begin{array}{cc}\begin{array}{c}\rho=\sum_{k=1}^Np_k|\phi_1^{(k)}\rangle\cdots|\phi^{(k)}_m\rangle\\
\langle\phi^{(k)}_m|\cdots\langle\phi^{(k)}_1|\end{array}&\left|\begin{array}{c}\left\||\phi^{(k)}_i\rangle\right\|^2=1,\;i=1,\ldots,m,\;k=1,\ldots,N,\\
\sum_{r,s=1}^Np_rp_s\prod_{i=1}^m\langle\phi^{(r)}_i|\phi^{(s)}_i\rangle\langle\phi^{(s)}_i|\phi^{(r)}_i\rangle=\|\varrho\|^2,\\
\sum_{k=1}^Np_k=1,\;p_k\geq 0,\;k=1,\ldots,N.\end{array}\right.\end{array}\right\},
\end{eqnarray*}
which is obviously bounded and closed. Since $\varrho$ a positive semidefinite $n\times n$ matrix, we can assume that $\varrho=\sum_{k=1}^K\alpha_k|\Psi^{(k)}\rangle\langle\Psi^{(k)}|$ be the orthogonal eigenvalue decomposition. Then, $\sum_{k=1}^K\alpha_k^2=\|\varrho\|^2$, and $\sum_{k=1}^K\alpha_k=1$ as $\mbox{Tr}(\varrho)=1$. Since $K\leq n$, we can find $\{|\phi^{(k)}_1\rangle,\ldots,|\phi^{(k)}_m\rangle\}_{k=1}^K$ such that
\begin{eqnarray*}
\begin{array}{c}
\left\||\phi^{(k)}_i\rangle\right\|^2=1,\;i=1,\ldots,m,\;k=1,\ldots,K,\\
\prod_{i=1}^m\langle\phi^{(r)}_i|\phi^{(s)}_i\rangle=0, \;\forall r\neq s,\;r,s=1,\ldots,K.
\end{array}
\end{eqnarray*}
Consequently, $\rho:=\sum_{k=1}^K\alpha_k|\phi_1^{(k)}\rangle\cdots|\phi^{(k)}_m\rangle\langle\phi^{(k)}_m|\cdots\langle\phi^{(k)}_1|\in S(\varrho)$. The result follows. \ep

The following proposition concerns some properties of the measure \reff{gm-m}.
\begin{Proposition}\label{prop-0}
Let $\varrho$ be the density matrix of a mixed state in ${\cal H}$ and $E(\varrho)$ be defined as \reff{gm-m}. Then, we have
\begin{itemize}
\item [(a)] $E(\varrho)\geq 0$ and $E(\varrho)=0$ if and only if $\varrho\in Disen({\cal H})$.
\item [(b)] Local unitary transformations on $Disen({\cal H})$ do not change $E$.
\end{itemize}
\end{Proposition}

\noindent {\bf Proof.} (a) By \reff{gm-m}, $E(\varrho)\geq 0$ for any $\varrho\in Disen({\cal H})$. If $\varrho\in Disen({\cal H})$, then with $\rho:=\varrho$, we get $\|\varrho-\rho\|=0$. consequently, $0\leq E(\varrho)\leq 0$ as desired. Now, suppose that $E(\varrho)=0$, i.e., there exists $\rho\in Disen({\cal H})$ such that $\|\varrho-\rho\|=0$. Consequently, $\varrho=\rho\in Disen({\cal H})$. The results follow.

(b) Denote by $\mathfrak{U}({\cal H})$ the group of local unitary linear transformations of ${\cal H}$. By the definition of $Disen({\cal H})$, it is obviously that $Disen({\cal H})$ is $\mathfrak{U}({\cal H})$-invariant. This, together with the fact that norm $\|\cdot\|$ is $\mathfrak{U}({\cal H})$-invariant, implies that $E$ is $\mathfrak{U}({\cal H})$-invariant. \ep


\section{The nearest disentangled mixed state}

\setcounter{Assumption}{0}
\setcounter{Theorem}{0} \setcounter{Proposition}{0}
\setcounter{Corollary}{0} \setcounter{Lemma}{0}
\setcounter{Definition}{0} \setcounter{Remark}{0}
\setcounter{Algorithm}{0}  \setcounter{Example}{0}
\hspace{4mm} In this section, we establish an analogue of Proposition \ref{prop-2} for mixed states based on the geometric measure defined by Definition \ref{def-3}. Like in \cite{wg03,q122}, where \reff{gm2} is considered instead of \reff{gm1}, we now, consider
\begin{eqnarray}\label{gm-m2}
\min_{\rho\in S(\varrho)}\frac{1}{2}\|\varrho-\rho\|^2
\end{eqnarray}
instead of \reff{gm-m}. Here $S(\varrho)$ is defined as that in Lemma \ref{lem-1}.
By the proof of Lemma \ref{lem-1}, the optimization problem \reff{gm-m2} can be parameterized as:
\begin{eqnarray}\label{gm-p}
\begin{array}{rcl}\min&&\displaystyle\frac{1}{2}\left\|\varrho-\sum_{k=1}^Np_k|\phi_1^{(k)}\rangle\cdots|\phi^{(k)}_m\rangle\langle\phi^{(k)}_m|\cdots\langle\phi^{(k)}_1|\right\|^2\\
\mbox{s.t.}&&\left\||\phi^{(k)}_i\rangle\right\|^2=1,\;i=1,\ldots,m,\;k=1,\ldots,N,\\
&&\sum_{r,s=1}^Np_rp_s\prod_{i=1}^m\langle\phi^{(r)}_i|\phi^{(s)}_i\rangle\langle\phi^{(s)}_i|\phi^{(r)}_i\rangle=\|\varrho\|^2,\\
&&\sum_{k=1}^Np_k=1,\;p_k\geq 0,\;k=1,\ldots,N.\end{array}
\end{eqnarray}
It is easy to see that \reff{gm-p} is equivalent to:
\begin{eqnarray}\label{gmm}
\begin{array}{rcl}\max&&\sum_{k=1}^Np_k\langle\phi^{(k)}_m|\cdots\langle\phi^{(k)}_1|\varrho|\phi_1^{(k)}\rangle\cdots|\phi^{(k)}_m\rangle\\
\mbox{s.t.}&&\left\||\phi^{(k)}_i\rangle\right\|^2=1,\;i=1,\ldots,m,\;k=1,\ldots,N,\\
&&\sum_{r,s=1}^Np_rp_s\prod_{i=1}^m\langle\phi^{(r)}_i|\phi^{(s)}_i\rangle\langle\phi^{(s)}_i|\phi^{(r)}_i\rangle=\|\varrho\|^2,\\
&&\sum_{k=1}^Np_k=1,\;p_k\geq 0,\;k=1,\ldots,N.\end{array}
\end{eqnarray}

\begin{Proposition}\label{prop-3}
The optimality conditions of maximization problem \reff{gmm} are:
\begin{eqnarray}\label{qe-m-1}
\left\{\begin{array}{l}p_k\langle\phi^{(k)}_m|\cdots\langle\phi^{(k)}_1|\varrho\prod_{j\neq i}|\phi_j^{(k)}\rangle=\mu_{ik}\langle\phi^{(k)}_i|\\
\;\;\;\;\;+\lambda p_k\sum_{t=1}^Np_t\left(\prod_{j\neq i}|\langle\phi^{(k)}_j|\phi^{(t)}_j\rangle|\right)^2(\langle\phi^{(k)}_i|\phi^{(t)}_i\rangle)\langle\phi^{(t)}_i|,\;i=1,\ldots,m,k=1,\ldots,N,\\
p_k\prod_{j\neq i}\langle\phi_j^{(k)}|\varrho|\phi_1^{(k)}\rangle\cdots|\phi^{(k)}_m\rangle=\mu_{ik}|\phi^{(k)}_i\rangle\\
\;\;\;\;\;+\lambda p_k \sum_{t=1}^Np_t\left(\prod_{j\neq i}|\langle\phi^{(k)}_j|\phi^{(t)}_j\rangle|\right)^2(\langle\phi^{(t)}_i|\phi^{(k)}_i\rangle)|\phi_i^{(t)}\rangle,\;i=1,\ldots,m,k=1,\ldots,N,\\
\langle\phi^{(k)}_m|\cdots\langle\phi^{(k)}_1
|\varrho|\phi_1^{(k)}\rangle\cdots|\phi^{(k)}_m\rangle=\lambda\sum_{t=1}^Np_t\left(\prod_{i=1}^m|\langle\phi^{(k)}_i|\phi^{(t)}_i\rangle|\right)^2+\kappa-\tau_k,\\
\;\;\;\;\;\;\;k=1,\ldots,N,\\
\tau_k,p_k\geq0,\tau_k p_k=0,\;k=1,\ldots,N,\\
\left\||\phi^{(k)}_i\rangle\right\|^2=1,\;i=1,\ldots,m,k=1,\ldots,N,\\
\sum_{r,s=1}^Np_rp_s\prod_{i=1}^m\langle\phi^{(r)}_i|\phi^{(s)}_i\rangle\langle\phi^{(s)}_i|\phi^{(r)}_i\rangle=\|\varrho\|^2,\\
\sum_{k=1}^Np_k=1.
\end{array}\right.
\end{eqnarray}
\end{Proposition}

\noindent {\bf Proof.} It follows from the Lagrange multiplier theorem and the concept of H-derivative in complex geometry \cite{h05}. \ep
\begin{Proposition}\label{prop-1}
Let $\varrho$ be the density matrix of a mixed state in ${\cal H}$ and $\{\lambda,p_k,\kappa,\tau_k,\mu_{ik},|\phi^{(k)}_i\rangle\}$ be a solution for \reff{qe-m-1}. We have the following conclusions.
\begin{itemize}
\item [(a)] $\mu_{1k}=\cdots=\mu_{mk}$ for any $k=1,\ldots,N$.
\item [(b)] Let $\mu_k:=\mu_{1k}=\cdots=\mu_{mk}$. Then, $\kappa=\sum_{k=1}^N\mu_k$.
\item [(c)] $\lambda\|\varrho\|^2+\kappa\in\mathbb{R}$ is a nonnegative real number and
\begin{eqnarray}\label{gm-c}
\sum_{k=1}^Np_k\langle\phi^{(k)}_m|\cdots\langle\phi^{(k)}_1|\varrho|\phi_1^{(k)}\rangle\cdots|\phi^{(k)}_m\rangle=\lambda\|\varrho\|^2+\sum_{k=1}^N\mu_k=\lambda\|\varrho\|^2+\kappa.
\end{eqnarray}
\end{itemize}
\end{Proposition}

\noindent {\bf Proof.} (a) By the first equation of \reff{qe-m-1}, we have that
\begin{eqnarray}\label{muk}
\mu_{ik}
&=&\left[p_k\langle\phi^{(k)}_m|\cdots\langle\phi^{(k)}_1|\varrho\prod_{j\neq i}|\phi_j^{(k)}\rangle-\lambda p_k\sum_{t=1}^Np_t\left(\prod_{j\neq i}|\langle\phi^{(k)}_j|\phi^{(t)}_j\rangle|\right)^2(\langle\phi^{(k)}_i|\phi^{(t)}_i\rangle)\langle\phi^{(t)}_i|\right]|\phi^{(k)}_i\rangle\nonumber\\
&=&p_k\left[\langle\phi^{(k)}_m|\cdots\langle\phi^{(k)}_1|\varrho|\phi_1^{(k)}\rangle\cdots|\phi_m^{(k)}\rangle-\lambda\sum_{t=1}^Np_t\left(\prod_{j=1}^m|\langle\phi^{(k)}_j|\phi^{(t)}_j\rangle|\right)^2\right],
\end{eqnarray}
which is independent of index $i$. Then, the result follows.

(b)
Let $\mu_k:=\mu_{1k}=\cdots=\mu_{mk}$. Multiplying the first equation of \reff{qe-m-1} by $|\phi^{(k)}_i\rangle$ and then subtracting $p_k$ times the third equation of \reff{qe-m-1}, we get
\begin{eqnarray*}
\mu_k=p_k\kappa-p_k\tau_k.
\end{eqnarray*}
This, together with the fourth and the last equations of \reff{qe-m-1}, implies that
\begin{eqnarray*}
\sum_{k=1}^N\mu_k=\kappa.
\end{eqnarray*}

(c) The result (b), together with the summation of the equations \reff{muk} from $k=1$ to $N$, implies \reff{gm-c}. Now, the facts that $p_k\geq 0$ and $\varrho$ is positive semidefinite imply that $\lambda\|\varrho\|^2+\kappa\in\mathbb{R}$ is a nonnegative real number. \ep

Similar to the entanglement eigenvalue for a pure state \reff{qe},
we define the entanglement eigenvalue for a mixed state.
\begin{Definition}\label{def-2}
Let $\varrho$ be the density matrix of a mixed state in ${\cal H}$.
\begin{eqnarray*}
\chi(\varrho):=\max\left\{\lambda\|\varrho\|^2+\kappa\;|\;\{\lambda,p_k,\tau_k,\kappa,\mu_{k},|\phi^{(k)}_i\rangle\}\; \rm satisfies\; \reff{qe-m-2}\right\}
\end{eqnarray*}
is called the entanglement eigenvalue of $\varrho$.
Here system \reff{qe-m-2} is defined as:
\begin{eqnarray}\label{qe-m-2}
\left\{\begin{array}{l}p_k\langle\phi^{(k)}_m|\cdots\langle\phi^{(k)}_1|\varrho\prod_{j\neq i}|\phi_j^{(k)}\rangle=\mu_{k}\langle\phi^{(k)}_i|\\
\;\;\;\;\;+\lambda p_k\sum_{t=1}^Np_t\left(\prod_{j\neq i}|\langle\phi^{(k)}_j|\phi^{(t)}_j\rangle|\right)^2(\langle\phi^{(k)}_i|\phi^{(t)}_i\rangle)\langle\phi^{(t)}_i|,\;i=1,\ldots,m,k=1,\ldots,N,\\
p_k\prod_{j\neq i}\langle\phi_j^{(k)}|\varrho|\phi_1^{(k)}\rangle\cdots|\phi^{(k)}_m\rangle=\mu_{k}|\phi^{(k)}_i\rangle\\
\;\;\;\;\;+\lambda p_k \sum_{t=1}^Np_t\left(\prod_{j\neq i}|\langle\phi^{(k)}_j|\phi^{(t)}_j\rangle|\right)^2(\langle\phi^{(t)}_i|\phi^{(k)}_i\rangle)|\phi_i^{(t)}\rangle,\;i=1,\ldots,m,k=1,\ldots,N,\\
\langle\phi^{(k)}_m|\cdots\langle\phi^{(k)}_1
|\varrho|\phi_1^{(k)}\rangle\cdots|\phi^{(k)}_m\rangle=\lambda\sum_{t=1}^Np_t\left(\prod_{i=1}^m|\langle\phi^{(k)}_i|\phi^{(t)}_i\rangle|\right)^2+\kappa-\tau_k,\\
\;\;\;\;\;\;\;k=1,\ldots,N,\\
\tau_k,p_k\geq0,\tau_k p_k=0,\;k=1,\ldots,N,\\
\left\||\phi^{(k)}_i\rangle\right\|^2=1,\;i=1,\ldots,m,k=1,\ldots,N,\\
\sum_{r,s=1}^Np_rp_s\prod_{i=1}^m\langle\phi^{(r)}_i|\phi^{(s)}_i\rangle\langle\phi^{(s)}_i|\phi^{(r)}_i\rangle=\|\varrho\|^2,\\
\sum_{k=1}^Np_k=1.
\end{array}\right.
\end{eqnarray}
\end{Definition}

Now, we have the following theorem.
\begin{Theorem}\label{thm-1}
Let $\varrho$ be the density matrix of a mixed state in ${\cal H}$. If $\chi(\varrho)$ is the entanglement eigenvalue of $\varrho$,
then
\begin{eqnarray}\label{sp-c}
\frac{1}{2}E(\varrho)^2=\|\varrho\|^2-\chi(\varrho).
\end{eqnarray}
Moreover, $\rho:=\sum_{k=1}^Np_k|\phi_1^{(k)}\rangle\cdots|\phi^{(k)}_m\rangle\langle\phi^{(k)}_m|\cdots\langle\phi^{(k)}_1|$ corresponding to $\chi(\varrho)$ is the nearest disentangled mixed state to $\varrho$.
\end{Theorem}

\noindent {\bf Proof.} It follows from \reff{gm-m2} and \reff{gmm}, Proposition \ref{prop-1} and Definitions \ref{def-3} and \ref{def-2} immediately. \ep

It is noted that $\chi(\varrho)$ is equal to the optimal value of problem \reff{gmm} and \reff{sp-c} reduces $1-\Lambda_{\max}$ for a pure state.

We now compute the geometric measure defined in \reff{gm-m2} for two examples. The computation is based on the maximization problem \reff{gmm}.

\begin{Example}\label{exm-1}
{\rm In this example, we consider the following bipartite qubit mixed state
\begin{eqnarray*}
\varrho&:=&\alpha\left(\frac{1}{\sqrt{2}}|00\rangle+\frac{1}{\sqrt{2}}|11\rangle\right)\left(\frac{1}{\sqrt{2}}\langle00|+\frac{1}{\sqrt{2}}\langle 11|\right)\nonumber\\
&&+(1-\alpha)\left(\frac{1}{\sqrt{2}}|01\rangle+\frac{1}{\sqrt{2}}|10\rangle\right)\left(\frac{1}{\sqrt{2}}\langle01|+\frac{1}{\sqrt{2}}\langle 10|\right)\nonumber,
\end{eqnarray*}
where $\alpha\in[0,1]$. It is easy to see $\frac{1}{2}E(\varrho)^2=\frac{1}{2}$ when both $\alpha=0$ and $\alpha=1$, which correspond to pure states. For general $\alpha\in(0,1)$, we use \reff{gmm} to compute $\frac{1}{2}E(\varrho)^2$.
It can be see that $n=4$ and $N=17$. Under the basis $\{|0\rangle,|1\rangle\}$, the corresponding maximization problem \reff{gmm} can be transformed into a maximization problem only involving real variables. By parameterizing $|\phi^{(k)}_j\rangle:=\left(\mathbf{x}^{(k,j)}_1+i\mathbf{y}^{(k,j)}_1\right)|0\rangle+\left(\mathbf{x}^{(k,j)}_2+i\mathbf{y}^{(k,j)}_2\right)|1\rangle$ for $j=1,2$ and $k=1,\ldots,17$,
we have
\begin{eqnarray}\label{gmmn}
\begin{array}{rcl}\max&&\sum_{k=1}^{17}p_k\left\{\frac{\alpha}{2}\left[\left(\left(\mathbf{x}^{(k,1)}\right)^T\mathbf{x}^{(k,2)}
-\left(\mathbf{y}^{(k,1)}\right)^T\mathbf{y}^{(k,2)}\right)^2\right.\right.\\
&&\;\;\;\;\;\;+\left.\left.\left(\left(\mathbf{x}^{(k,1)}\right)^T\mathbf{y}^{(k,2)}+\left(\mathbf{y}^{(k,1)}\right)^T\mathbf{x}^{(k,2)}\right)^2\right]\right.\\
&&+\left.\frac{1-\alpha}{2}\left[\left(\mathbf{x}^{(k,1)}_2\mathbf{x}^{(k,2)}_1+\mathbf{x}^{(k,1)}_1\mathbf{x}^{(k,2)}_2
-\mathbf{y}^{(k,1)}_1\mathbf{y}^{(k,2)}_2-\mathbf{y}^{(k,1)}_2\mathbf{y}^{(k,2)}_1\right)^2\right.\right.\\
&&\;\;\;\;\;\;+\left.\left.\left(\mathbf{y}^{(k,1)}_1\mathbf{x}^{(k,2)}_2+\mathbf{y}^{(k,2)}_2\mathbf{x}^{(k,1)}_1
+\mathbf{y}^{(k,2)}_1\mathbf{x}^{(k,1)}_2+\mathbf{y}^{(k,1)}_2\mathbf{x}^{(k,2)}_1\right)^2\right]\right\}\\
\mbox{s.t.}&&\left(\mathbf{x}^{(k,1)}\right)^T\mathbf{x}^{(k,1)}+\left(\mathbf{y}^{(k,1)}\right)^T\mathbf{y}^{(k,1)}=1,\;k=1,\ldots,17,\\
&&\left(\mathbf{x}^{(k,2)}\right)^T\mathbf{x}^{(k,2)}+\left(\mathbf{y}^{(k,2)}\right)^T\mathbf{y}^{(k,2)}=1,\;k=1,\ldots,17,\\
&&\sum_{r,s=1}^{17}p_rp_s\prod_{i=1}^2\left\{\left[\left(\mathbf{x}^{(r,i)}\right)^T\mathbf{x}^{(s,i)}+\left(\mathbf{y}^{(r,i)}\right)^T\mathbf{y}^{(s,i)}\right]^2\right.\\
&&\;\;\;\;\;+\left.\left[\left(\mathbf{x}^{(r,i)}\right)^T\mathbf{y}^{(s,i)}-\left(\mathbf{x}^{(s,i)}\right)^T\mathbf{y}^{(r,i)}\right]^2\right\}=1+2\alpha^2-2\alpha,\\
&&\sum_{k=1}^{17}p_k=1,\;p_k\geq 0,\;k=1,\ldots,17.\end{array}
\end{eqnarray}
Problem \reff{gmmn} is solved using MatLab Optimization ToolBox, which can always find a good local maximizer. The result is shown in Figure 1.
For $\alpha=0.5$, the nearest $\rho$ computed is
\begin{eqnarray*}
\rho=\sum_{k=1}^4p_k |\phi^{(k)}_1\rangle|\phi^{(k)}_2\rangle\langle\phi^{(k)}_2|\langle\phi^{(k)}_1|
\end{eqnarray*}
with $|\phi^{(k)}_j\rangle:=\left(\mathbf{x}^{(k,j)}_1+i\mathbf{y}^{(k,j)}_1\right)|0\rangle+\left(\mathbf{x}^{(k,j)}_2+i\mathbf{y}^{(k,j)}_2\right)|1\rangle$ for $j=1,2$ and the parameters being in Table 1.
\begin{table}[h]\label{tab1}
  \caption{Parameters for the nearest disentangled mixed state}
  \begin{center}
  \tabcolsep 2.8pt
  \renewcommand\arraystretch{1.0}
      \begin{tabular}[c]{ c || c | c | c | c | c |c | c | c | c }
      \hline \hline
                         k&          $p_k$&      $\mathbf{x}^{(k,1)}_1$ &     $\mathbf{x}^{(k,1)}_2$ &   $\mathbf{x}^{(k,2)}_1$ &     $\mathbf{x}^{(k,2)}_2$ & $\mathbf{y}^{(k,1)}_1$ &     $\mathbf{y}^{(k,1)}_2$ & $\mathbf{y}^{(k,2)}_1$ &     $\mathbf{y}^{(k,2)}_2$           \\ \hline \hline
                         1&0.0414  &  0.4495   & 0.4497    &0.6749     &  0.6747   &  0.5458  &  0.5457   & 0.2113   &  0.2114\\ \hline
                         2&0.2163  &   0.5211  &   0.5210  &  0.1227   & 0.1227    & 0.4780   &  0.4781   & 0.6963   & 0.6964 \\ \hline
                         3&0.5000  &  0.5572   &  -0.5572  & -0.7061   &   0.7061  & -0.4353  &  0.4353   &  0.0369  & -0.0369 \\ \hline
                         4& 0.2423 &   0.6908  &  0.6909   & 0.3020    &0.3020     & 0.1506   & 0.1508    & 0.6393   & 0.6395 \\ \hline \hline
    \end{tabular}
  \end{center}
 \end{table}

}
\end{Example}

We now consider a class of bipartite qubit mixed states with less symmetric structures.
\begin{Example}\label{exm-2}
{\rm In this example, we consider the following bipartite qubit mixed state
\begin{eqnarray*}
\varrho&:=&\alpha\left(\gamma_1|00\rangle+\gamma_2|11\rangle\right)\left(\gamma_1\langle00|+\gamma_2\langle 11|\right)\nonumber\\
&&+(1-\alpha)\left(\gamma_3|01\rangle+\gamma_4|10\rangle\right)\left(\gamma_3\langle01|+\gamma_4\langle 10|\right)\nonumber,
\end{eqnarray*}
where $\alpha\in[0,1]$, $\gamma_1^2+\gamma_2^2=1$ and $\gamma_3^2+\gamma_4^2=1$. The optimization problem is similar to \reff{gmmn}. For {\bf Case I}: $\gamma_1=\gamma_3:=\frac{1}{\sqrt{3}}$ and $\gamma_2=\gamma_4:=\sqrt{\frac{2}{3}}$, and {\bf Case II}: $\gamma_1:=\frac{1}{\sqrt{3}}$ and $\gamma_2:=\sqrt{\frac{2}{3}}$, and $\gamma_3:=\frac{1}{\sqrt{4}}$ and $\gamma_4:=\sqrt{\frac{3}{4}}$, the computational results are shown in Figure 2. We see that the curve of {\bf Case II} is not symmetric with respect to $\alpha=0.5$, which agrees with the choice of parameters. The other cases for parameters $\alpha,\gamma$ have similar phenomena.
}
\end{Example}


\begin{figure}[htbp]
\centering
\includegraphics[width=5.6in]{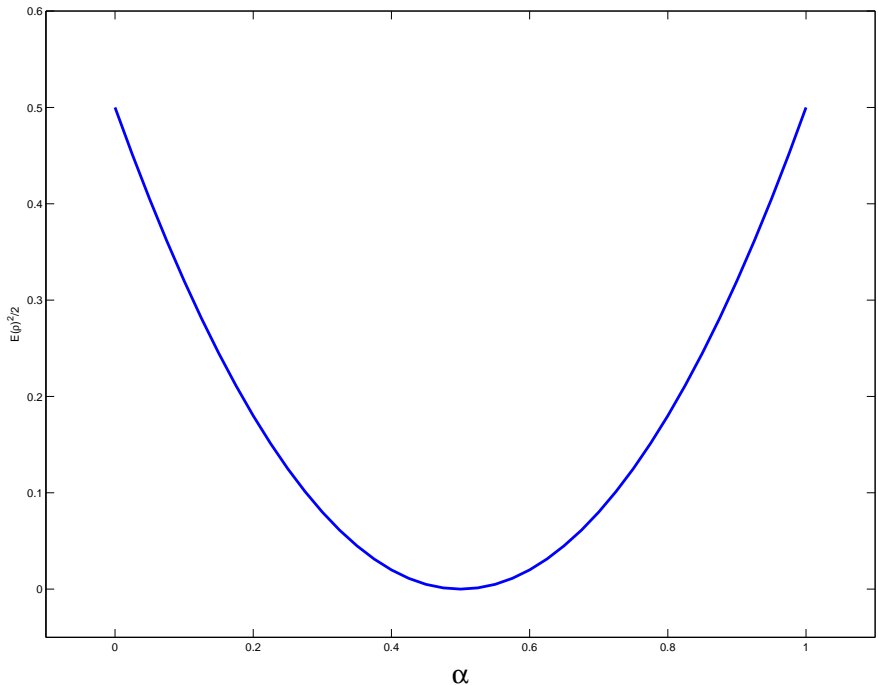}
\caption{The measure $\frac{E(\varrho)^2}{2}$ for the mixed states in Example \ref{exm-1}}
\end{figure}

\begin{figure}[htbp]
\centering
\includegraphics[width=5.6in]{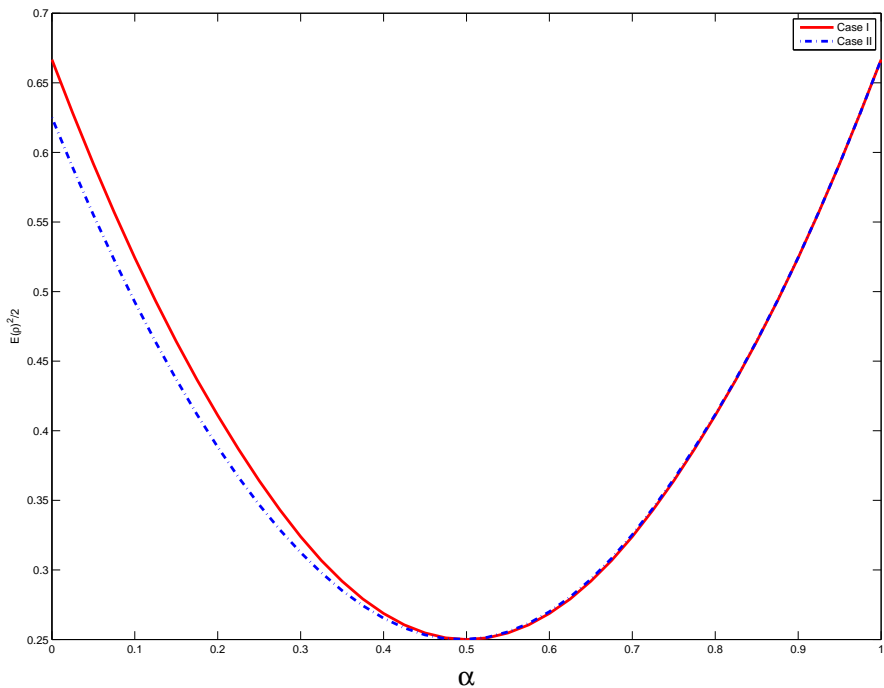}
\caption{The measure $\frac{E(\varrho)^2}{2}$ for the mixed states in Example \ref{exm-2}}
\end{figure}

\section{Conclusion}

\setcounter{Assumption}{0}
\setcounter{Theorem}{0} \setcounter{Proposition}{0}
\setcounter{Corollary}{0} \setcounter{Lemma}{0}
\setcounter{Definition}{0} \setcounter{Remark}{0}
\setcounter{Algorithm}{0}  \setcounter{Example}{0}

\hspace{4mm} We have extended the geometric measure to mixed states and
established a characterization of the nearest disentangled mixed
state of a given mixed state with respect to this measure. The
analogue results for the quantum eigenvalue of a pure state are
established for mixed states, i.e., Proposition \ref{prop-1} and
Theorem \ref{thm-1}. Based on this geometric measure, further works
on the analysis and the computation are desired.


\end{document}